\documentclass[runningheads]{llncs}
\usepackage{xcolor}
\usepackage{multirow}
\usepackage{graphicx}
\usepackage{soul}
\usepackage{float}
\begin{document}
%
\title{Harmonization Across Imaging Locations (HAIL): One-Shot Learning for Brain MRI}
\titlerunning{Harmonization Across Imaging Location(HAIL)}
%
\author{Abhijeet Parida\inst{1} \and Zhifan Jiang \inst{1}  \and Syed Muhammad Anwar\inst{1, 4} \and Nicholas Foreman\inst{3} \and Nicholas Stence\inst{3} \and Michael J. Fisher\inst{2} \and Roger J. Packer\inst{1} \and Robert A. Avery\inst{2} \and Marius George Linguraru\inst{1, 4}}

%
\authorrunning{A. Parida et al.}
%
\institute{Children’s National Hospital, Washington, DC, USA \and Children’s Hospital of Philadelphia, Philadelphia, PA, USA \and Children’s Hospital Colorado, Aurora, CO, USA \and  George Washington University, Washington, DC, USA}
%
\maketitle              
\begin{abstract}
For machine learning-based prognosis and diagnosis of rare diseases, such as pediatric brain tumors, it is necessary to gather medical imaging data from multiple clinical sites that may use different devices and protocols. Deep learning-driven harmonization of radiologic images relies on generative adversarial networks (GANs). However, GANs notoriously generate pseudo structures that do not exist in the original training data, a phenomenon known as ”hallucination”. To prevent hallucination in medical imaging, such as magnetic resonance images (MRI) of the brain, we propose a one-shot learning method where we utilize neural style transfer for harmonization. 
At test time, the method uses one image from a clinical site to generate an image that matches the intensity scale of the collaborating sites.
Our approach combines learning a feature extractor, neural style transfer, and adaptive instance normalization. We further propose a novel strategy to evaluate the effectiveness of image harmonization approaches with evaluation metrics that both measure image style harmonization and assess the preservation of anatomical structures. Experimental results demonstrate the effectiveness of our method in preserving patient anatomy while adjusting the image intensities to a new clinical site. Our general harmonization model can be used on unseen data from new sites, making it a valuable tool for real-world medical applications and clinical trials. 

\end{abstract}

\section{Introduction}
\label{sec:intro} 

Deep learning (DL)-based models trained on large radiologic data with high-quality labels are effective for clinical diagnosis and trials. However, to achieve clinically useful outcomes for rare diseases such as pediatric brain tumors, data collection requires collaboration between multiple clinical centers. 
Only then, the amount of data generally required to effectively train such models could be made available. Since clinical centers use different imaging equipment and often varying acquisition protocols, we are presented with significant challenges for the analysis and interpretation of radiological imaging data such as magnetic resonance imaging (MRI). Since there is no underlying standardized unit in MRIs, they may have different intensities and anatomical resolutions. Further, MRIs are subject to domain shifts arising from a wide range of scanning parameters and differences in populations across clinical centers. Such domain shifts between training and testing data (e.g. new unseen site) could lead to increased errors in clinical tasks performed using machine learning algorithms \cite{guan2023domainatm,torralba2011unbiased,wilson2020survey}. Therefore, multi-site data must be pre-processed with harmonization to obtain a uniform appearance and allow machine learning algorithms to be effectively trained~\cite{pomponio2020harmonization}. However, such intensity harmonizations could adversely affect anatomical information in a scan, if not properly managed.  

The diversity in medical imaging data poses challenges to traditional but limited-intensity harmonization methods, such as histogram matching~\cite{nyul_hm_2000,shah_hm_2011}. Deep learning approaches that map an image from a source to a target domains have the additional benefit of combining spatial and anatomical features information to achieve intensity harmonization. These methods typically rely on types of generative adversarial networks (GANs)~\cite{goodfellow2016nips}, such as conditional GANs~\cite{mirza2014conditional} that translate between domains using paired images. However, it is rare to find medical images from the same patient acquired at multiple sites. Alternatives like CycleGAN ~\cite{mirza2014conditional} learn two GANs by enforcing cycle consistency, thus forgoing the need for paired data. In addition, unsupervised image-to-image translation (UNIT)~\cite{liu_unit_2017} combines GANs with a variational autoencoder~\cite{kingma2019introduction} and uses a shared latent space for harmonization. The UNIT model has been applied to MRI data to generate a harmonized optimal domain, but exclusively for segmentation \cite{carlos}. Unfortunately, GAN-based methods do not enforce structural consistency to preserve patient anatomy during image transformation. Conserving patient anatomy is paramount for accurate diagnosis and treatment. Without structural consistency, the generated images could lose clinically relevant details~\cite{yang_cyclegan_2018}.

Therefore, we focus on intensity harmonization for MRI data, while preserving patient-specific anatomical information. The first inspiration for our work is the neural style transfer (NST), a technique that uses neural networks to generate images by combining the anatomy of a input image and the intensity of a target image~\cite{adain}. The main assumption is that the patient anatomy in an MRI scan remains the same, regardless of the imaging site~\cite{medstyle}. The differences in MRI appearances is due to changes in scanners or protocols. An adaptive instance normalization (AdaIN) module~\cite{adain} aligns the distribution of the anatomical features with that of the target features to achieve harmonized features. For 2D image harmonization, NST employ pre-trained VGG models \cite{vgg} as feature extractors. The advantages of using such a pre-trained model diminishes when the new task deviates from the task for which the model was initially trained \cite{koch2015siamese}. Therefore, for an optimal MRI feature extractor,  we need to train a 3D feature extractor specific to the downstream data. NST methods, such as ~\cite{medstyle,unc,Torbati_2021_ICCV}, jointly minimize two losses for the prediction- content loss from the input image and the style loss from the target image. We design a NST framework to handle 3D data using the 3D feature extractor and AdaIN to minimize style and content losses.


The second inspiration for the study is one-shot learning technique which learns from a limited set of data, making it a valuable tool for 
rare diseases \cite{meta_me}. While few-shot learning for image-to-image translation has been used in image registration \cite{he2021few}, for image harmonization, we must translate the intensity while preserving the anatomy. One training strategy used for one-shot learning is called meta-learning \cite{finn2017model,ravi2017optimization}, which learns a model in two stages- an unrelated training stage(meta-learning phase) and a task-specific learning stage \cite{KHADKA2022105227,meta_me}. Convolutional Siamese networks are common one-shot learning architectures \cite{koch2015siamese},  which have branched networks to learn highly discriminative representations of the inputs, even with limited training data \cite{sung2018learning}. Branched networks can predict outcomes on unseen data by enforcing similarity at test time, which is an advantage for medical imaging tasks. 

\begin{figure*}[!hb]
  \centerline{\includegraphics[width=0.9\linewidth]{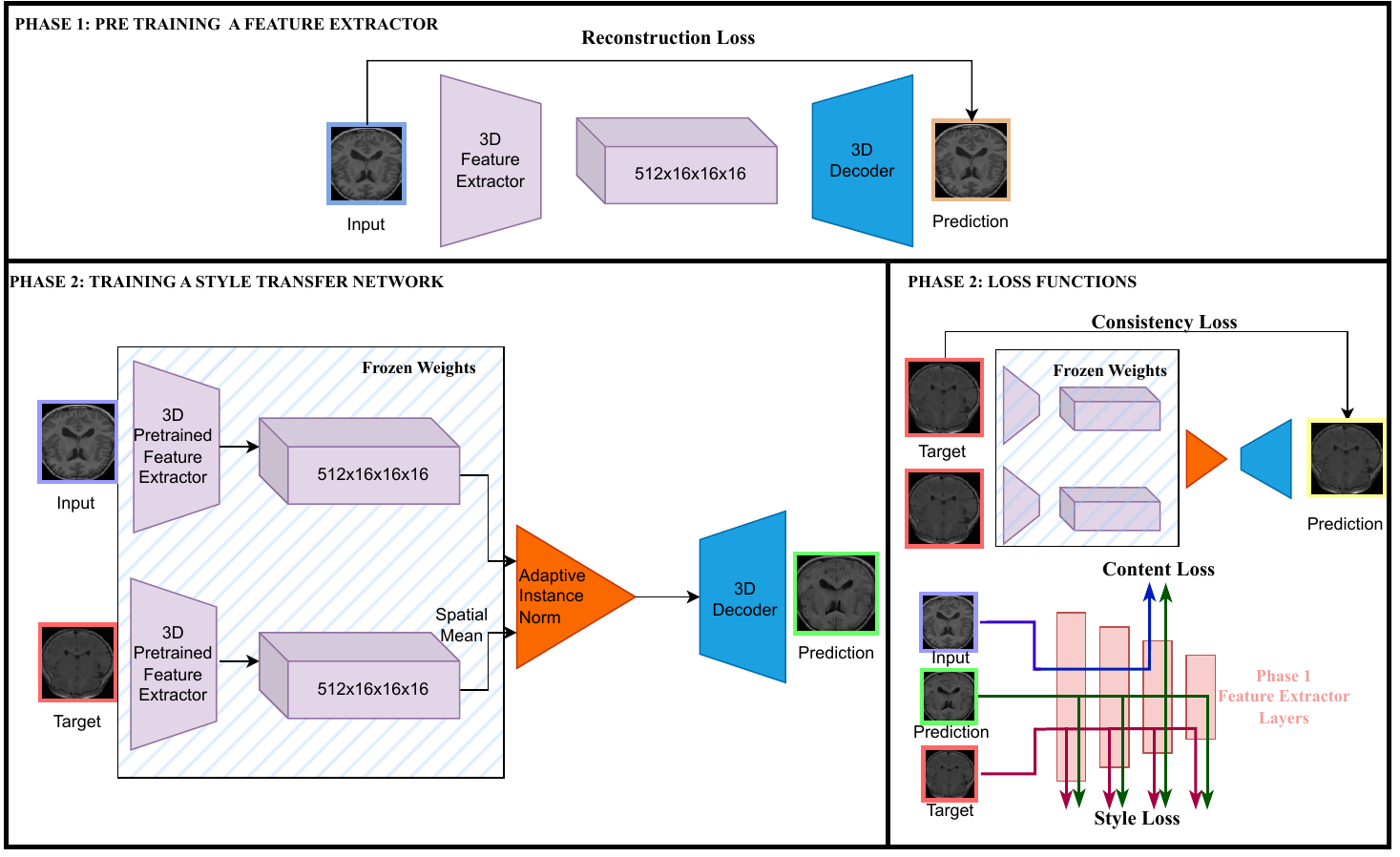}}
\caption{\textbf{Harmonization across imaging locations (HAIL) framework.} The input and target MRIs each pass through a 3D feature extractor to produce latent representations. These representations are then passed through a 3D adaptive instance normalization (AdaIN) module, which translates them for the decoder to produce the predicted image-harmonized MRI. The loss function includes a consistency loss, which serves as a regularizer to prevent over-correction during image harmonization. The style loss and content loss are calculated based on features extracted by the layers of the pre-trained 3D feature extractor.}
\label{fig:schema}
\end{figure*}

To address these requirements for medical image harmonization, we propose the harmonization across imaging locations (HAIL) framework illustrated in Figure \ref{fig:schema}. Our novel method has four major contributions: 

\begin{enumerate}
    \item Novel modular NST framework that harmonizes 3D medical images. 
    \item One-shot learning image harmonization framework that learns broad features, thus generalizing to data from unseen test sites using one target sample. 
    \item Novel metrics for measuring intensity harmonization and preservation of anatomical structures to allow future methods to be fairly compared.
    \item Evaluation of the effectiveness of the proposed approach for harmonizing multi-site MRI data from rare diseases, i.e., pediatric brain tumors. 
\end{enumerate}
\section{Method \& Experimental Setup}

\subsection{Image Harmonization} \label{sec:dt}
The proposed framework- HAIL, using one-shot learning has two phases- 1) a feature extractor for meta-learning and 2) learning a task-specific 3D NST model with AdaIN \cite{adain}. We use four different losses for model training- reconstruction loss, consistency loss, style loss, and content loss. The reconstruction loss is used to train the 3D feature extractor. The content loss ensures similarity in activation of the higher layers for input and predicted images \cite{adain}. Whereas, style loss ensures similar feature statistics for the prediction and target images \cite{adain}. We introduce the notion of consistency loss to the loss landscape to prevent harmonization when target and input images are similar.   

\subsubsection{Phase 1: Pre-training a feature extractor.}
In this meta-learning phase, we trained an encoder-decoder architecture (Appendix \ref{app:arch}) to compress and reconstruct an image (Figure \ref{fig:schema}). The training is governed by reconstruction loss and in the process, a latent space is generated which is used by the decoder for image reconstruction. Later, we froze the encoder parameters and used them to extract features and compute the content and style losses in phase 2.

\noindent\textbf{Implementation details:}
Images from all three sites A, B, and C were divided into training and validation sets- using 80:20 splits. The input image was 3D cropped to $64\times  64 \times  64$ patch. The algorithms were implemented on the lightning \cite{PyTorch} framework and trained on an NVIDIA RTX A5000 using half-precision (FP16). The encoder-decoder was optimized to minimize the reconstruction loss using $AdamW$ optimizer, batch size $48$, and learning rate $1e^{-4}$. The reconstruction loss was a combination of L1 and structural similarity (SSIM) \cite{MONAI} losses with equal weights. The model was trained for 1,000 epochs and the best validation model was saved for phase 2. 

\subsubsection{Phase 2: Training a style transfer model for one-shot learning.}
To learn the task-specific and dataset agnostic style transfer between the 3D images, we used a Convolutional Siamese network (Figure \ref{fig:schema}). The twin network with identical  weights reused the frozen encoder from phase 1 to extract the input and target image features. The target image acted as the single example for the one-shot image harmonization. The input features are translated to the target site using AdaIN \cite{adain}. The decoder, with same architecture from phase 1, takes the translated features and generates a stylized image corresponding to the intensity harmonized image.

\noindent\textbf{Implementation details:} 
The images from sites A and B were divided into training, validation, and testing sets using 70:20:10 split. Site C was reserved to test the generalizability of the HAIL framework in one-shot learning.  Each instance in a batch has a pair of images- input, and target. The input image was 3D cropped $64\times  64 \times  64$ patch. The paired target image patch was created by cropping the corresponding location of the target image. An instance of the batch used site A image as input and site B image as target. The next instance used the site B image as input and the site A image as target. So, we learned harmonization between sites (A  $\rightarrow$ B and B $\rightarrow$ A) simultaneously. This combined with the random sampling of image pairs for training helps prevent overfitting and train one-shot learners. The decoder was optimized to minimize the content \cite{adain}, style \cite{adain}, and consistency loss function using AdamW optimizer with initial learning rate $1e^{-4}$ and batch size $32$. The learning rate decayed by 0.8 when validation loss plateaued. The consistency loss was a combination of the L1 and SSIM losses with equal weights. The weights ($\lambda$) between style, content, and consistency losses were $\lambda_{style}=100$, $\lambda_{content}=150$, and $\lambda_{consistency}=200$, respectively. The choice of the weights was made to bring the losses in the order of magnitude of $10^{-1}$. The model was trained for 1,000 epochs and the best validation model was saved to report metrics. 

\subsection{Image Harmonization Evaluation}\label{me}
The evaluation strategy assess 1) intensity harmonization, i.e., the appearance of the predicted image match that of the target image, and 2) anatomy preservation, i.e., the structures in the input image are preserved even after harmonization. To this end, we propose using Wasserstein distance (WD)~\cite{wasser} to evaluate intensity harmonization by measuring the movement of intensity histograms. We chose WD over Jenson-Shannon (JS) or Kullback-Leibler (KL) divergences, since JS divergence is a fixed value for non-overlapping distributions, and KL divergence is not defined for non-overlapping distributions \cite{kolouri2018sliced}. 
We define $WD(i,t)$ as WD between input ($i$) and target ($t$) images as the upper bound for the model prediction performance. To make the metric agnostic to the magnitude of scales for different sites and make it comparable between sites, we report the normalized WD defined as 
\begin{equation}
 nWD(i,p)\%=\frac{WD(i,p)}{WD(i,t)} \times 100 \quad \textrm{and} \quad nWD(t,p)\%=\frac{WD(t,p)}{WD(i,t)} \times 100,   
\end{equation}
where $WD(i,p)$ is the WD between input($i$) and prediction ($p$) and $WD(t,p)$ between target ($t$) and prediction ($p$). For good performance in intensity harmonization, we expect a large $nWD(i, p)$ and a small $nWD(t, p)$.

To evaluate anatomy preservation, we propose using a method that automatically segments anatomical structures in the input and the predicted image for comparison. This also checks if the output is suitable to be used for a downstream DL-based task, such as segmentation. Since minor changes in clinical information maybe critical, we propose using relative absolute volume difference (rAVD) for comparing the segmentation results. 
\begin{equation}
 rAVD\%= \frac{|vol(p)-vol(i)|}{vol(i)} \times 100,   
\end{equation}
where $vol(i)$ and $vol(p)$ denote input and prediction volumes for a structure. For good performance in anatomy preservation, we expect a small $rAVD$.

\noindent\textbf{Implementation details:} 
For calculating $nWD$ for harmonizing from A $\rightarrow$ B, we pick one sample from site B (example for one-shot learning) as target and made prediction on test samples of site A as input. To segment anatomy, we used a robust model from Freesurfer v7~\cite{fischl2012freesurfer} to segment the brain gray matter (GM) and white matter (WM). Freesurfer models have been trained on large datasets and are robust to a wide range of data shifts. Also, most importantly they are publicly accessible and noted as an acceptable performance by the community.   

\section{Results} \label{sec:dis}
\subsection{Data and Pre-processing}
We collected full head MRIs of pediatric brain tumor patients from three clinical sites: A, B, and C. Each site provided $n = 60$ 3D T1-weighted MRIs using different scanners and acquisition protocols (details in Table ~\ref{tab:data}). We applied N4 bias field correction and using an MRI from site A as reference performed inter-subject rigid registration using advanced normalization tools (ANTs)~\cite{ants_itk_2014}.

Due to computational resource limitation and the fact that we focus only on intensity harmonization, MRI resolution was changed to $1\times1\times1 \;mm^3$ and was resized to $256\times256\times256$ voxels. All voxels were re-scaled to $[0, 1]$ using the min-max normalization. 
The inverse transforms were stored to convert the images back to values that are clinically meaningful. 
\begin{table}[h]
\caption{\textbf{Dataset summary} displays the acquisition protocols for pediatric brain MRIs at each site. }
\resizebox{1\linewidth}{!}{
\begin{tabular}{l|c|c|c|c|c|c}

\hline
       & MANUFACTURER     & \begin{tabular}[c]{@{}c@{}}ACQUISITION\\ PLANE\end{tabular} & \begin{tabular}[c]{@{}c@{}}ECHO \\ TIME ($ms$) \end{tabular}& \begin{tabular}[c]{@{}c@{}} REPETITION \\ TIME ($ms$) \end{tabular} & \begin{tabular}[c]{@{}c@{}}IN-PLANE  ($mm^2$)\\ RESOLUTION\end{tabular} & \begin{tabular}[c]{@{}l@{}}SLICE ($mm$)\\ THICKNESS\end{tabular} \\ \hline
SITE A & General Electric & Axial                                                       & 10.5   & 600    & $0.41\times0.41$                                                                              & 0.6                                                            \\
SITE B & Siemens          & Sagittal                                                    & 2.5    & 1900   & $0.82\times0.82$                                                                             & 0.9                                                            \\
SITE C & Phillips         & Coronal                                                     & 3.8    & 8.23      & $0.94\times0.94$                                                                              & 1.0                                                            \\ \hline
\end{tabular}
}
\label{tab:data}
\end{table}
\subsection{Image Harmonization}
\subsubsection*{Intensity harmonization:}
\begin{figure*}[ht]
  \centerline{\includegraphics[width=1\linewidth]{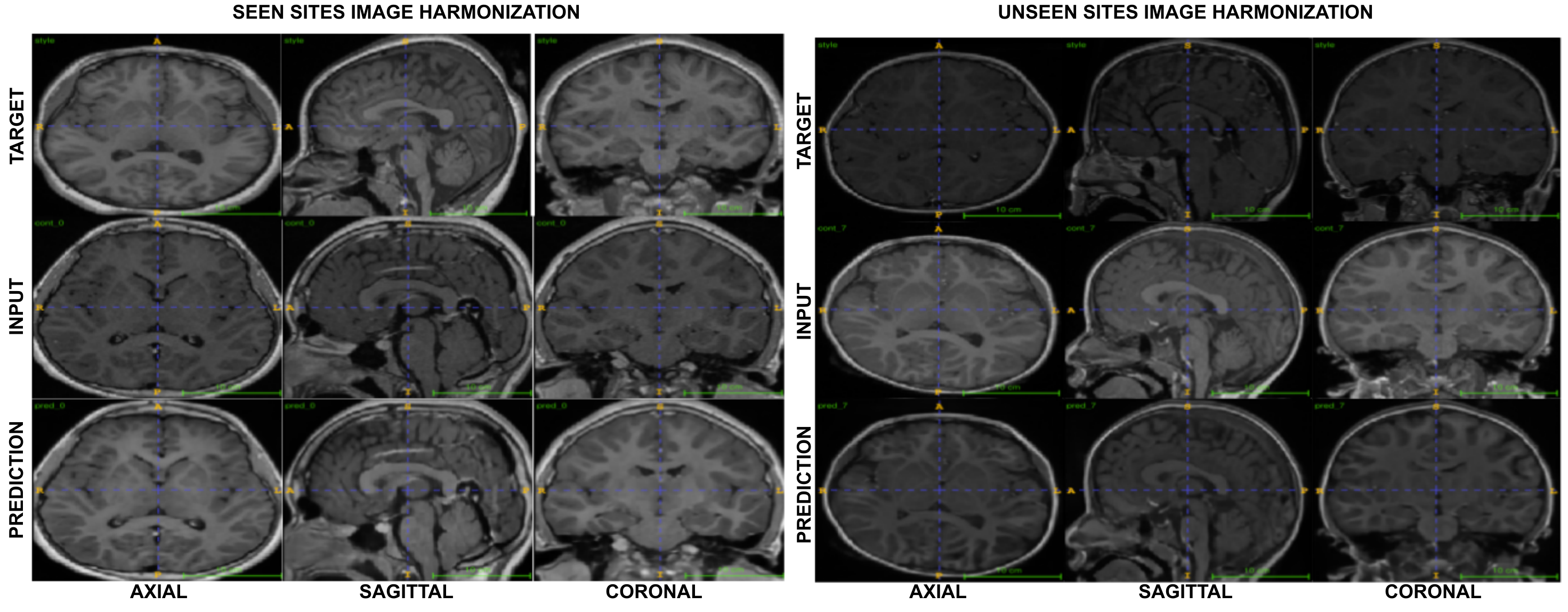}}
\caption{\textbf{Qualitative results of image harmonization.}  We show axial, sagittal, and coronal slices of the 3D input, target, and predicted MRIs. The predicted MRI preserved the anatomical structures from the input MRI, while the intensities are aligned with that of the target MRI. The image shows good harmonization of the model for data from both seen and unseen sites. 
}
\label{fig:results}
\end{figure*}

As shown in Table \ref{tab:quant-res}, HAIL achieves a higher $nWD(i,p)$ (average=0.94) compared to $nWD(t,p)$ (average=0.11), so the prediction has moved away from the input intensity domain and is closer to the target. This is visually confirmed in Fig. \ref{fig:results}, where the predicted intensity resembles the target intensity. Further, $nWD(t,p)$ is low for both seen and unseen sites, indicating that HAIL is not specific to the style transfer A $\rightarrow$ B $\rightarrow$ A, but can be used for transfers between any pairs of sites. To test this outcome, we added data from an independent site C, and used a single target image to demonstrate generalizability of the one-shot harmonization strategy (Table \ref{tab:quant-res} unseen sites).

\subsubsection*{Anatomy Preservation:}
Visual inspection of Fig. \ref{fig:results} for both seen and unseen sites shows the input and prediction have similar shapes, sizes and structures. Quantitatively, as seen in Table \ref{tab:quant-res}, the perceived anatomical change due to harmonization is rAVD = 7.06\% for GM and rAVD = 13.42\% for WM. Thus, HAIL preserves the anatomy well within the clinically acceptable margin of error.
\begin{table*}[hb]
\caption{\textbf{Quantitative results} for image harmonization calculated for various sites, the metrics are presented as avg$\pm$std across all test samples in the dataset. Higher $nWD(i, p)$ compared to $nWD(t, p)$ indicates good harmonization of intensities, while low  $rAVD$ means anatomies are preserved during the harmonization. "$\star$" shows significant($p<=0.05$) performance differences between HAIL and Liu et al. method \cite{medstyle} using the Wilcoxon signed-rank test.}
\resizebox{1\linewidth}{!}{
\centering
{
\begin{tabular}{cl|cc|cc|cc|cc}
\hline
\multicolumn{2}{c|}{\multirow{2}{*}{Sites}} & \multicolumn{2}{c|}{nWD(i,p) $\%$} & \multicolumn{2}{c|}{nWD(t,p) $\%$} & \multicolumn{2}{c|}{rAVD(GM) $\%$} & \multicolumn{2}{c}{rAVD(WM) $\%$} \\ \cline{3-10} 
\multicolumn{2}{c|}{} & HAIL & Liu et. al.\cite{medstyle} & HAIL & Liu et. al\cite{medstyle} & HAIL & Liu et. al\cite{medstyle} & HAIL & Liu et. al\cite{medstyle} \\ \hline
\multirow{3}{*}{\rotatebox[origin=c]{90}{SEEN}} & A $\rightarrow$ B & \textbf{92.27±2.03} & 90.63±3.88 & 15.71±0.31 & \textbf{12.76±0.63}$^\star$ & \textbf{6.99±16.76} & 12.05±19.31 & \textbf{18.86±35.28} & 43.94±26.64$^\star$ \\
 & B $\rightarrow$ A & 96.16±2.01 & \textbf{96.75±3.31} & 9.04±0.21 & \textbf{7.47±0.68}$^\star$& 6.78±4.71 & \textbf{6.27±4.71}& \textbf{7.69±8.47} & 21.40±12.70$^\star$ \\ \cline{2-10} 
 & avg & \textbf{94.22} & 93.69 & 12.38 & \textbf{10.12} & \textbf{6.89} & 9.16 & \textbf{13.28} & 32.67 \\ \hline
\multirow{5}{*}{\rotatebox[origin=c]{90}{UNSEEN}} & A $\rightarrow$ C & \textbf{94.81±2.28} & 73.04±4.22$^\star$ & \textbf{9.43±0.15} & 27.31±0.43$^\star$& \textbf{12.50±12.55} & 21.99±20.47$^\star$ & \textbf{19.97±19.58} & 68.45±46.36$^\star$ \\
 & C $\rightarrow$ A & \textbf{97.99 ±2.69} & 85.83±5.13$^\star$ & 13.87±0.18 & \textbf{11.64±0.75} & \textbf{4.16±2.77} & 4.17±4.67 & \textbf{9.03±5.90} & 17.07±19.92$^\star$ \\
 & B $\rightarrow$ C & \textbf{94.59±2.25} & 85.11±3.60$^\star$ & \textbf{7.12±0.14} & 15.45±0.36$^\star$& 9.73±9.32 & \textbf{6.79±7.73} & 21.79±15.45 & \textbf{18.83±13.07}\\
 & C $\rightarrow$ B & \textbf{91.92±3.92} & 88.90±7.36$^\star$ & \textbf{16.06±0.17} & 19.21±0.53$^\star$& \textbf{2.17±2.16} & 3.86±3.56& \textbf{3.16±2.92} & 8.58±22.92$^\star$ \\ \cline{2-10} 
 & avg & \textbf{94.83} & 83.22 & \textbf{11.62} & 18.40 & \textbf{7.14} & 9.20 & \textbf{13.49} & 28.23 \\ \hline
\multicolumn{2}{c|}{Overall} & \textbf{94.63} & 88.46 & \textbf{11.87} & 14.26 & \textbf{7.06} & 9.18 & \textbf{13.42} & 30.45 \\ \hline
\end{tabular}
}
}
\label{tab:quant-res}
\end{table*}

\subsection{Comparison with State-of-the-Art}
We compared the performance of HAIL with a GAN-based NST approach \cite{medstyle}, 
which harmonized 2D images and aggregated them to generate a 3D output. 
As shown in Table \ref{tab:quant-res}, 
for the seen sites, the model in \cite{medstyle} performs similar for $nWD(i,p)$ and better for $nWD(i,t)$ by $\sim2\%$ when compared with HAIL. However, HAIL strategy achieved better $rAVD$, which is clinically a meaningful metric. Further, for the unseen sites, we had two observations. First, the GAN-based model failed to converge and produce meaningful output for two samples during the A $\rightarrow$ C harmonization, while HAIL converged on all data. Second, HAIL significantly outperformed the approach in \cite{medstyle} on $nW D(i,p), nW D(t,p)$ and $rAVD(WM)$($p<=0.05$) by an average margin of 11\%, 7\%, and 16\% respectively. The performance was similar for $rAVD(GM)$, where the improvement was 2\%. This suggests that HAIL generalizes better than the GAN-based model 
when learning from a small training dataset and with the addition of a new site.  

\subsection{Impact of Consistency Loss}
We hypothesized that consistency loss in HAIL  acts as a regularizer and aids towards better image harmonization performance in a one-shot manner. To investigate this, the model was retrained using the exact same parameters and seeds but with $\lambda_{consistency} = 0$.
The model performance for intensity harmonization trained with consistency loss, was lower for the seen sites in terms of $nWD$($p<=0.05$), as seen in Table \ref{tab:quant-consis}. However, the performance with consistency loss was better for unseen sites by a margin of 9\% for $nWD(i, p)$ and 10\% for $nWD(t, p)$($p<=0.05$ for both). The consistency loss model performed better for both- seen data (2\% for $rAVD(GM)$ and 6\% for $rAVD(WM$), $p<=0.05$ for both), and unseen data (6\% for $rAVD(GM)$ and 8\% for $rAVD(WM)$, $p<=0.05$ for both). These findings have implications for the design and optimization of image harmonization models, as they demonstrate that incorporating consistency loss is important for generalizability for unseen sites. 
\begin{table*}
\caption{\textbf{Impact of consistency loss} on image harmonization, the metrics are presented as avg$\pm$std across all test samples in the dataset.  "$\star$" shows significant ($p<=0.05$) performance differences between HAIL with and without the consistency loss using the Wilcoxon signed-rank test.
}
\resizebox{1\linewidth}{!}{
\centering
{
\begin{tabular}{cl|cc|cc|cc|cc}
\hline
\multicolumn{2}{c|}{\multirow{2}{*}{Sites}} & \multicolumn{2}{c|}{nWD(i,p) $\%$} & \multicolumn{2}{c|}{nWD(t,p) $\%$} & \multicolumn{2}{c|}{rAVD(GM) $\%$} & \multicolumn{2}{c}{rAVD(WM) $\%$} \\ \cline{3-10} 
\multicolumn{2}{c|}{} & with loss & without loss & with loss & without loss &  with loss & without loss &  with loss & without loss \\ \hline
\multirow{3}{*}{\rotatebox[origin=c]{90}{SEEN}} & A $\rightarrow$ B & 92.27±2.03 & \textbf{96.32±1.91}$^\star$ & 15.71±0.31 & \textbf{9.58±0.16}$^\star$ & \textbf{6.99±16.76} & 8.18±14.49$^\star$ & \textbf{18.86±35.28} & 23.53±54.64 \\
 & B $\rightarrow$ A & \textbf{96.16±2.01} & 96.09±1.91 & 9.04±0.21 & \textbf{7.24±0.12}$^\star$& \textbf{6.78±4.71} & 8.35±5.71$^\star$& \textbf{7.69±8.47} & 15.42±10.51$^\star$ \\ \cline{2-10} 
 & avg & 94.22 & \textbf{96.21} & 12.38 & \textbf{8.41} & \textbf{6.89} & 8.27 & \textbf{13.28} & 19.47 \\ \hline
 
\multirow{5}{*}{\rotatebox[origin=c]{90}{UNSEEN}} & A $\rightarrow$ C & \textbf{94.81±2.28} & 68.07±1.96$^\star$ & \textbf{9.43±0.15} & 35.41±0.23$^\star$& \textbf{12.50±12.55} & 28.20±14.36$^\star$ & \textbf{19.97±19.58} & 34.54±20.01$^\star$ \\
 & C $\rightarrow$ A & 97.99 ±2.69 & \textbf{99.83±2.64} & 13.87±0.18 & \textbf{12.22±0.13} & \textbf{4.16±2.77} & 4.93±3.68 & \textbf{9.03±5.90} & 15.07±9.92$^\star$ \\
 & B $\rightarrow$ C & \textbf{94.59±2.25} & 79.69±1.83$^\star$ & \textbf{7.12±0.14} & 25.51±0.19$^\star$& \textbf{9.73±9.32} & 17.15±10.36$^\star$ & \textbf{21.79±15.45} & 29.32±18.67$^\star$\\
 & C $\rightarrow$ B & 91.92±3.92 & \textbf{95.25±3.92}$^\star$ & 16.06±0.17 & \textbf{14.47±0.19}$^\star$& \textbf{2.17±2.16} & 2.22±2.13& \textbf{3.16±2.92} & 5.04±5.29$^\star$ \\ \cline{2-10} 
 & avg & \textbf{94.83} & 85.71 & \textbf{11.62} & 21.90 & \textbf{7.14} & 13.13 & \textbf{13.49} & 21.18 \\ \hline
\multicolumn{2}{c|}{Overall} & \textbf{94.63} & 90.96 & \textbf{11.87} & 15.16 & \textbf{7.06} & 10.70 & \textbf{13.42} & 20.33 \\ \hline
\end{tabular}
}
}
\label{tab:quant-consis}
\end{table*}

\section{Conclusion}
\label{sec:conclu}
Rare diseases present unique challenges for clinical trial design and implementation due to limited data availability. In our study, we suggest using a deep learning framework for image harmonization (HAIL) can improve the quality of multi-site data and increase the statistical power of analyses. We showed how a neural style transfer model can achieve good intensity harmonization for 3D medical scans by learning generic features, which allows training generic image harmonization models. These methods are one-shot learners as they can adapt an input image to a target intensity domain by using only one image from unseen data at test time. We also proposed metrics that would allow future methods for  medical image harmonization to be fairly compared. Our results demonstrated that HAIL improved the consistency of multi-site, multi-protocol data and could lead to better generalizability of deep learning models. 

\newpage
\section{Acknowledgments}
\label{sec:acknowledgments}
This work was possible due to the support from the National Cancer Institute (Grant No: UG3CA236536) and US Department of Defense (Grant No : W81XWH1910376). 
\bibliographystyle{splncs04}
\bibliography{refs}
\newpage
\appendix
\section{Network Architectures}\label{app:arch}
\begin{figure*}[h]
\centerline{\includegraphics[width=0.9\linewidth]{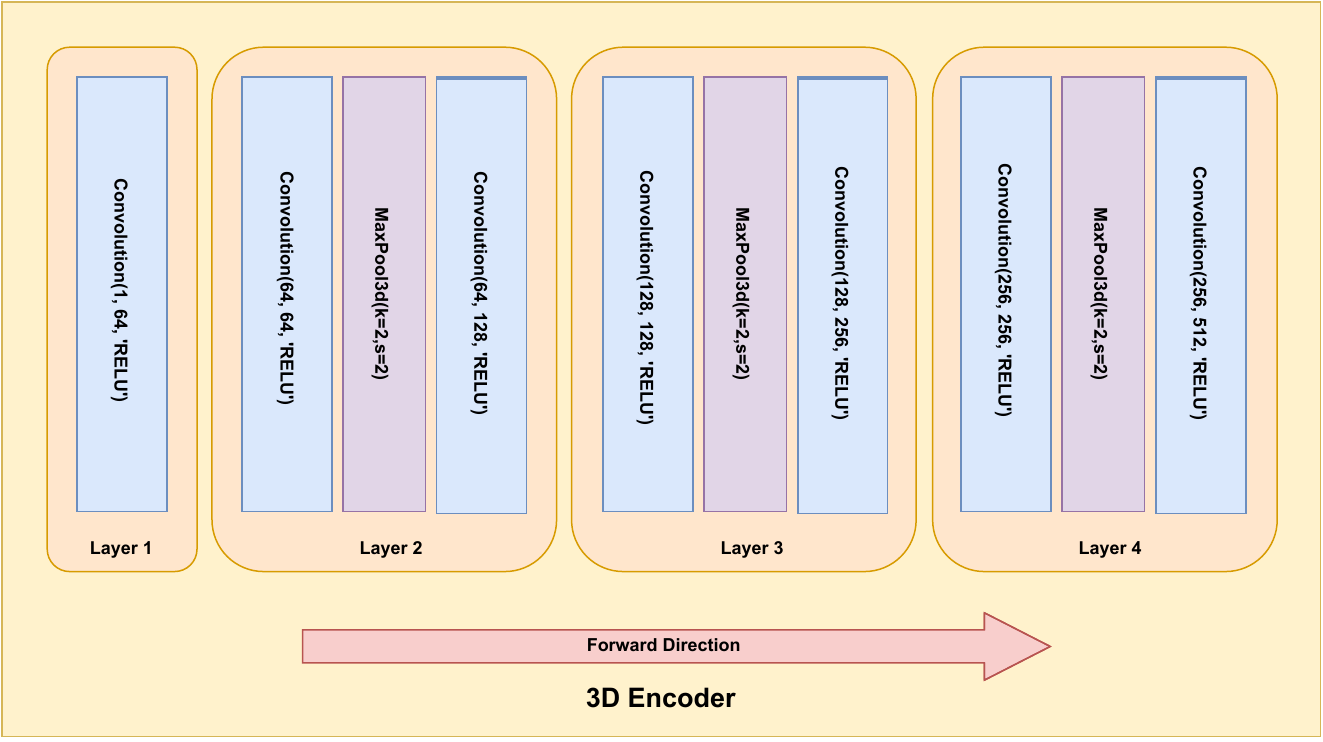}}
\caption{\textbf{Encoder architecture.} The image shows the various layers of the encoder architecture for the proposed HAIL framework. Convolution(.) refers to the Convolution implementation in monai.networks.blocks. Convolution(1, 64, 'RELU') means it is a convolution layer with $spatial\_dims=3$, $in\_channels=1$, $out\_channels=64$, $kernel\_size=3$, $stride=1$, $padding=1$, followed by a $ReLU$ non-linearity and normalization as $None$. MaxPool3d(.) refers to the MaxPool3d implementation in torch.nn. MaxPool3d(k=2, s=2) means a 3D max pooling operation with $kernel\_size=2$ and $stride=2$. The features from layer 4 for the input and target are passed into the AdaIN module. Each layer of the encoder is used to extract features for the calculation of the style and content losses.}
\label{fig:schema-encoder}
\end{figure*}
\begin{figure*}[h]
\centerline{\includegraphics[width=1\linewidth]{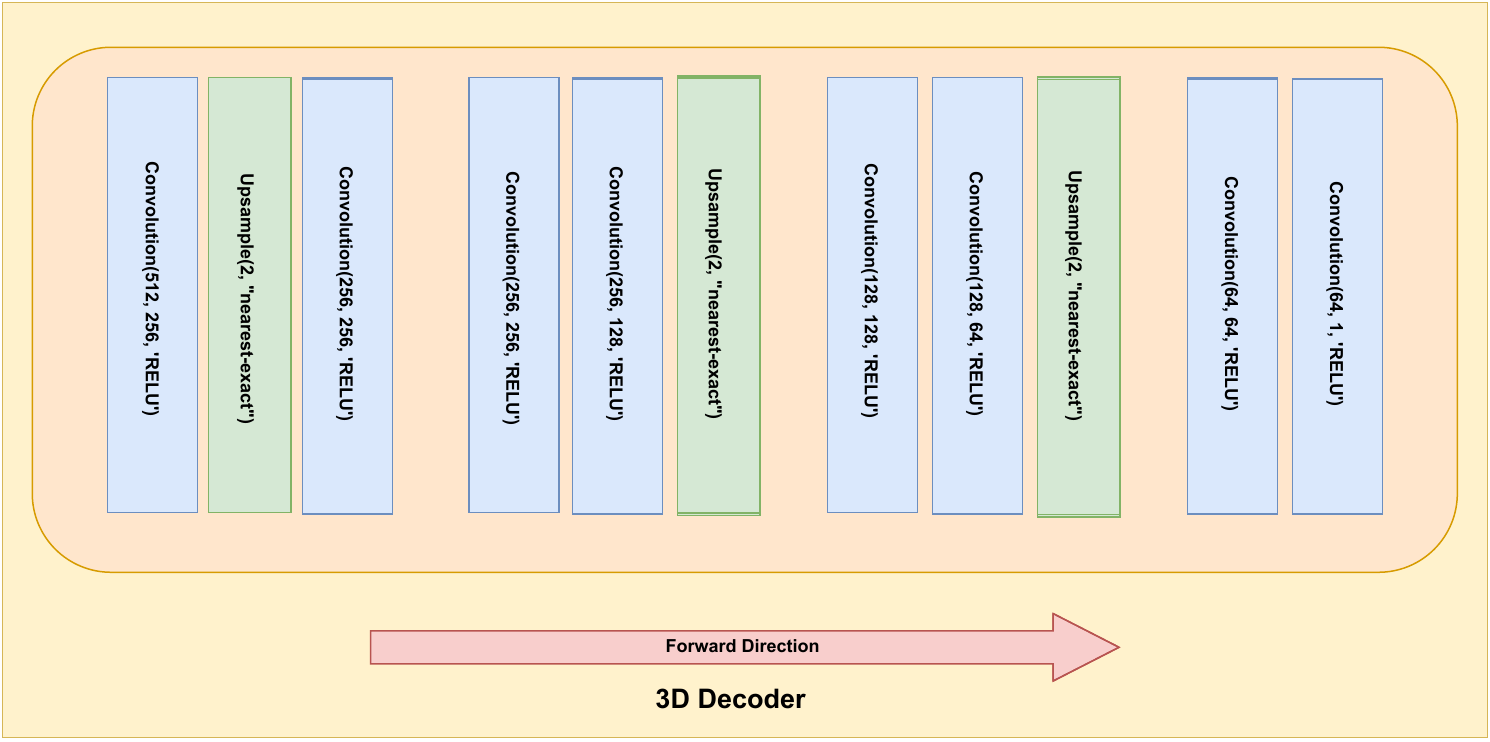}}
\caption{\textbf{Decoder architecture.} The image shows the various layers of the decoder used to generate a 3D brain MRI from the 3D harmonized features from the AdaIn module. for the proposed HAIL framework. Convolution(.) refers to the Convolution implementation in monai.networks.blocks. Convolution(512, 256, 'RELU') means it is a convolution layer with $spatial\_dims=3$, $in\_channels=512$, $out\_channels=256$, $kernel\_size=3$, $stride=1$, $padding=1$, followed by a $ReLU$ non-linearity and normalization as $None$. MaxPool3d(.) refers to the Upsample implementation in torch.nn. Upsample(2, 'nearest-exact') means a 3D upsampling operation with $scale\_factor=(2, 2, 2)$ and $mode='nearest-exact'$.}
\label{fig:schema-decoder}
\end{figure*}
\end{document}